\documentclass{appolb}
\usepackage{graphicx}
\usepackage[numbers,sort&compress]{natbib}
\usepackage{amsmath}
\usepackage{enumerate}
\usepackage{tikz}
\usepackage{wrapfig}
\usepackage{mathtools}
\usepackage{nicefrac}
\graphicspath{{images/}}
\usepackage{amssymb}
\usepackage{amsfonts}
\usepackage{physics}
\usepackage{xspace}
\usepackage{bm}
\usepackage{mathrsfs}
\usepackage{nicefrac}
\usepackage{url}
\usepackage{hyperref}
\usepackage{doi}
\usepackage{array}

\newcommand{\beq}{\begin{eqnarray}}
\newcommand{\eeq}{\end{eqnarray}}

\newcommand{\bqa}{\begin{eqnarray}}
\newcommand{\eqa}{\end{eqnarray}}

\newcommand{\bea}{\begin{eqnarray}}
\newcommand{\eea}{\end{eqnarray}}

\newcommand{\be}{\begin{eqnarray*}}
\newcommand{\ee}{\end{eqnarray*}}

\newcommand{\bel}[1]{\begin{eqnarray}\label{#1}}
\newcommand{\eel}{\end{eqnarray}}

\newcommand{\nn}{\nonumber}

\newcommand{\EQ}[1]{Eq.~(\ref{#1})}
\newcommand{\EQn}[1]{(\ref{#1})}

\newcommand{\EQSn}[1]{(\ref{#1})}

\newcommand{\EQSTWO}[2]{Eqs.~(\ref{#1})~and~(\ref{#2})}
\newcommand{\EQSTWOn}[2]{(\ref{#1})~and~(\ref{#2})}

\newcommand{\CITn}[1]{\cite{#1}}

\newcommand{\av}{{\boldsymbol a}} 
\newcommand{\bv}{{\boldsymbol b}} 
\newcommand{\evs}{{\boldsymbol e}}

\newcommand{\vv}{{\boldsymbol v}}
\newcommand{\pvs}{{\boldsymbol p}}

\newcommand\alphav{{\boldsymbol \alpha}}

\newcommand\varepsv{{\boldsymbol \varepsilon}}

\newcommand{\trthree}{{\rm tr_3}}

\definecolor{amethyst}{rgb}{0.6, 0.4, 0.8}

\date{\today}

\begin{document}
\title{Spin alignment, tensor polarizabilities, and local equilibrium for spin-1 particles}

\author{
Wojciech Florkowski, Sudip Kumar Kar, Valeriya Mykhaylova
\address{Institute of Theoretical Physics, Jagiellonian University, 30-348 Krak\'ow, Poland}
}
\maketitle
\begin{abstract}
Different bases for the spin-1 density matrix are discussed to clarify the connection between its components and observables measured in heavy-ion collisions. The theoretical advantage of using the adjoint representation for spin matrices is emphasized. Next, the equilibrium spin density matrix and the corresponding Wigner function are introduced. With appropriate definitions of the energy-momentum and spin tensors, this framework allows for the formulation of perfect spin hydrodynamics in the same way as previously done for spin-\nicefrac{1}{2} particles. Together, these results provide a unified description of spin-\nicefrac{1}{2} and spin-1 particles.
\end{abstract}


\section{Introduction}

The measurements of spin-related observables in heavy-ion collisions have introduced a new perspective for studies of the properties of strongly interacting matter. In addition to the data on particle abundances, spectra, and correlations, we have gained a completely new characteristic of matter. It is often emphasized that, through studies of spin polarization, we enter a genuinely quantum realm of particle production processes. 

The experimental measurements of spin polarization can be divided into two categories: measurements of the spin polarization of spin-$\nicefrac{1}{2}$ hadrons, mainly $\Lambda$ hyperons~\CITn{STAR:2017ckg, Adam:2018ivw, Niida:2018hfw, STAR:2018gyt, STAR:2019erd, STAR:2025jhp, STAR:2025jwc}, and the measurements of tensor polarizabilities of spin-1 hadrons, mainly vector mesons~\CITn{STAR:2008lcm, ALICE:2019aid,  STAR:2022fan, Shen:2024rdq}. In the last case, one concentrates on the coefficient $\rho_{00}$ or its deviation from 1/3 called alignment. On the theoretical side, these measurements have been analyzed (and in some cases originally proposed) in \CITn{Becattini:2021iol, Palermo:2024tza, Banerjee:2024xnd} and \CITn{Liang:2004xn, Yang:2017sdk, Sheng:2019kmk, Xia:2020tyd, Sheng:2020ghv, Muller:2021hpe, Li:2022vmb, Kumar:2023ojl, Yin:2025gvl,Li:2025pef}, respectively, for hyperon and vector-meson results. For reviews, see, for example \CITn{Florkowski:2018fap, Mohanty:2021vbt, Becattini:2024uha}. Theoretical advances directly related to spin physics include studies of the Wigner functions~\CITn{Becattini:2013fla, Florkowski:2018ahw, Weickgenannt:2019dks, Hattori:2020gqh, Huang:2020kik} as well as the development of spin hydrodynamics~\CITn{Florkowski:2017ruc, Weickgenannt:2020aaf, Weickgenannt:2021cuo, Bhadury:2020puc, Shi:2020htn, Hattori:2019lfp, Hu:2021pwh, Fang:2025aig, Fukushima:2020ucl, Montenegro:2017rbu}.

The briefly sketched landscape of spin physics above highlights the importance of a deeper investigation of the relationships between different observables, as well as between theory and experiment. In this paper, we study fundamental relations among spin observables, aiming to clarify several issues discussed in recent literature.

We start our discussion with the spin-1 density matrix $\rho$ and argue that its theoretical description is most conveniently formulated in the adjoint representation, in which the antisymmetric part of $\rho$ is directly expressed by the spin polarization vector ${\cal P}^\mu$, while the symmetric part (after subtracting the Kronecker delta contribution) is given by the tensor polarizabilities ${\cal T}^{\mu\nu}$. This representation is connected with the use of longitudinal-polarization vectors $\epsilon^\mu$ which are reduced to unit real vectors in the particle rest frame~(PRF). The adjoint representation proves particularly useful in calculating the Wigner function for a relativistic gas of spin-1 particles, which can then be employed to obtain the energy-momentum and spin tensors. 

The central part of this work is devoted to the analysis of {\it local equilibrium}. We argue that it can be defined analogously to the concept introduced for spin-$\nicefrac{1}{2}$ particles, i.e., as a state in which the spin part of the total angular momentum is conserved separately. The adopted form of the local-equilibrium spin density matrix uniquely determines the spin polarization vector and the tensor polarizabilities. This result is particularly important, since it allows one to establish a connection between the hyperon and vector-meson measurements. Previously, such a connection was found in the coalescence model~\CITn{Xia:2020tyd}. Here we employ a different physical picture, in which hadrons, rather than quarks, are equilibrated. A unified description of spin-1 and spin-$\nicefrac{1}{2}$ particles allows the construction of perfect spin hydrodynamics that simultaneously incorporates both types of particles. 

Finally, we turn to the discussion of tensor polarizabilities. Since they are measured in the basis in which the operator $J_y$ is diagonal, it is necessary  to clarify the relation between this basis and the adjoint representation. We find that the experimentally measured alignment is directly expressed by the coefficient ${\cal T}^{22}$ evaluated in the PRF.

\smallskip 
{\it Conventions and notation:} For the Levi-Civita symbol $\epsilon^{\mu\nu\alpha\beta}$, we follow the convention \mbox{$\epsilon^{0123} =-\epsilon_{0123} = +1$}. The tensor  \mbox{${\tilde a}^{\alpha\beta} \equiv (\nicefrac{1}{2}) \,\epsilon^{\alpha\beta\gamma\delta} a_{\gamma \delta}$} is dual to $a^{\alpha\beta}$. The metric tensor is of the form \mbox{$g_{\mu\nu} = \textrm{diag}(+1,-1,-1,-1)$}. Throughout the text, we make use of natural units, \mbox{$\hbar = c = k_{\rm B} = 1$}. The scalar product of two four-vectors $a$ and $b$ reads $a \cdot b = a^0 b^0 - \av \cdot \bv$, where three-vectors are indicated in bold. The symbol $\tr$ denotes the trace over Lorentz indices, for example, $\tr(\rho) = \rho^\mu_{\,\,\mu}$, while the trace over spin indices is denoted by $\trthree$, e.g., $\trthree(\rho^A) = \Sigma_{r=1}^3\, \rho^A_{rr}$. The antisymmetric part of a tensor $T^{\mu\nu}$ is defined as $T^{[\mu\nu]} = (T^{\mu\nu}-T^{\nu\mu})/2$.
\section{Spin density matrix for spin-1 particles}

\begin{figure}[t]
    \centering
    \includegraphics[width=0.3\linewidth]{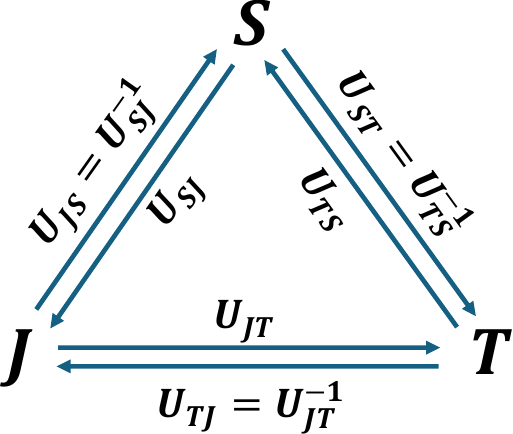}
    \caption{Schematic diagram of the transformations between different representations, defined by the unitary matrices $U_{AB}$, whose action is given by $U^\dagger_{AB} A U_{AB} = B$, where $A,B = J,S,T$.}
    \label{fig:JTS_tr}
\end{figure}

Information about the spin state is fully encoded in the spin density matrix~\cite{Leader:2001nas}. However, its explicit form depends on the specific choice of both the Lorentz reference frame and the quantum basis used for the spin states. For a massive particle, the spin density is commonly defined in the PRF obtained by the canonical boost\footnote{The boost directly determined by the three-velocity of the particle; for its explicit form see, for example, \CITn{Florkowski:2017dyn}.}. Regarding the spin-1 basis, three options are commonly used. In the first (standard) case, we employ the operators $J_x, J_y, J_z$, which satisfy the angular-momentum algebra, with $J_z$ being diagonal. In the second case, the spin operators are chosen as $T_x = J_y, T_y = J_z$ and $T_z = J_x$ (i.e.~with diagonal $T_y$), while the third case considers the adjoint representation with operators \mbox{$S_x = S^1, S_y = S^2, S_z = S^3$}, defined by the relation
\begin{equation}
(S^i)_{jk} = -i \epsilon_{ijk}.
\label{eq:Sijk}
\end{equation}
Operators $J, T$ and $S$ are equivalent and are related by similarity transformations of the form
\begin{equation}
U^\dagger_{AB} A U_{AB} = B, \quad A,B = J,S,T,
\quad U^\dagger_{AB} U_{AB} = 1,
\label{eq:UAB}
\end{equation}
presented schematically in Fig.~\ref{fig:JTS_tr}. 

We note that, in general, the boost changes the directions of the reference frame axes, which in turn affects the choice of the spin basis~\cite{Florkowski:2021pkp,Florkowski:2023zms}. However, this effect was found to be very small in current heavy-ion collisions~\cite{STAR:2022fan}, therefore we neglect it in the present discussion.

The use of a basis with diagonal $J_z$ seems to be natural when considering the spin polarizations of both vector mesons and $\Lambda$ hyperons. In the latter case, we always use the Pauli matrices with diagonal $\sigma_z$. However, spin-1-related measurements determine the spin alignment in the basis where $T_y$ is diagonal~\cite{STAR:2022fan}. The adjoint representation arises naturally if we use the states corresponding to linear polarization~\cite{Leader:2001nas}. We thus see that different bases serve different purposes, making it useful to establish simple relations between them. 

In the PRF (denoted by an asterisk\footnote{We note that the asterisk (appearing as an upper or lower index) should not be confused with the complex conjugate.}), the spin density matrix can be written in the form~\cite{Leader:2001nas}
\begin{equation}
\rho^S_* = \frac{1}{3} \left[ 
1 + \frac{3}{2} {\cal P}^i_* S^i + \sqrt{\frac{3}{2}} {\cal T}^{ij}_* \left( 
S^i S^j + S^j S^i \right)
\right],
\label{eq:rhoS}
\end{equation}
where ${\cal P}^i_*$ are the components of the spin polarization vector and ${\cal T}^{ij}_*$ are tensor polarizabilities. In Eq.~(\ref{eq:rhoS}), we employed the spin operators $S^i$ but one can equally well use $J^i$ or $T^i$. This yields alternative expressions, namely $\rho^J$ and  $\rho^T$, that are interconnected by the unitary transformation~(\ref{eq:UAB}). Using (\ref{eq:Sijk}) in (\ref{eq:rhoS}) we find
\begin{align}
    \rho^S_{rs*}  &= \frac{1}{3} \left[\delta_{rs} + \frac{3}{2} {\cal P}^i_* (S^i)_{rs} + \sqrt{\frac{3}{2}}{\cal T}^{ij}_* \left(S^i S^j + S^j S^i \right)_{rs}\right]
    \nonumber \\
    &=\frac{1}{3} \left[\delta_{rs} - \frac{3i}{2} {\cal P}^i_* \epsilon_{irs} - \sqrt{6}\, {\cal T}_{rs*}\right],
\end{align}
where $r,\,s=1,2,3$ are the spin indices.

It should be emphasized that the spin polarization vector ${\cal P}^i_*$ and the tensor polarizabilities ${\cal T}^{ij}_*$ are independent of the choice of representation and are determined by the traces~\cite{Leader:2001nas}:
\begin{eqnarray}
{\cal P}^i_* &=& \trthree \left(\rho^A_* A^i \right), \quad A=J,\,S,\,T,  \\
{\cal T}^{ij}_* &=& \frac{1}{2}\sqrt{\frac{3}{2}}\left(\trthree\left[\rho^A_* (A^i A^j + A^j A^i)\right] -\frac{4}{3}\delta^{ij}\right).
\end{eqnarray}
Note that the element $\mathcal{T}^{22}_*$, which we find to be directly related to spin alignment (see~Sec.~\ref{sec:alignent}), may be written as
\begin{align}
    \mathcal{T}^{22}_* &=\frac{1}{\sqrt{6}}\left(\tr_3 \left[ \rho_*^A (A^2)^2\right]  -\tr_3 \left[\rho_*^A (A^1)^2 \right] \right. \nonumber \\
  &  \left.
   +\tr_3\left[\rho_*^A (A^2)^2\right]-\tr_3\left[\rho_*^A (A^3)^2\right]\right).
\label{eq:aniso}   
\end{align}
Analogous expressions can also be found for $ \mathcal{T}^{11}_*$ and $ \mathcal{T}^{33}_*$, indicating that the components $\mathcal{T}^{ii}_*$ are sensitive to the anisotropy of the spin densities. 

\section{J and S representations}

In this section, we collect the most important information about the J and S representations. Although the formulas presented here are well known and can be found in most of the quantum mechanics textbooks, assembling them together helps us to proceed with the analysis of spin polarization effects for spin-1 particles. The explicit forms of the $J^i$ spin matrices are: 

 \begin{align}
 \hspace{-0.3cm} J_x\!=\! \frac{1}{\sqrt{2}}\!\left(
\begin{array}{ccc}
 \!0 & \!1 & \!0 \\
 \!1 & \!0 & \!1 \\
 \!0 & \!1 & \!0 \\
\end{array}
\right)\!, \,
J_y \!=\! \frac{i}{\sqrt{2}}\!\left(
\begin{array}{ccc}
 \!0 & \!-1 & \!0 \\
 \!1 & \!0 & \!-1 \\
 \!0 & \!1 & \!0 \\
\end{array}
\right)\!,\,
J_z\!=\!\!\left(
\begin{array}{ccc}
 \!1 & \!0 & \!0 \\
 \!0 & \!0 & \!0 \\
 \!0 & \!0 & \!-1 \\
\end{array}
\right).
\label{eq:J}
\end{align}
The basis used in (\ref{eq:J}) consists of the eigenvectors of the matrix $J_z$, commonly denoted as $|1,+1 \rangle$, $|1,0 \rangle$, and $|1,-1 \rangle$. For brevity of notation, we will also refer to these states as $|1\rangle = |1,+1 \rangle$, $|2\rangle =|1,0 \rangle$, and $|3\rangle = |1,-1 \rangle$. The spin matrices in the adjoint representation are defined by Eq.~(\ref{eq:Sijk}), which gives
\begin{align}
S_x=\!\left(
\begin{array}{ccc}
 0 & 0 & 0 \\
 0 & 0 & -i \\
 0 & i & 0 \\
\end{array}
\right), \,
S_y=\!\left(
\begin{array}{ccc}
 0 & 0 & i \\
 0 & 0 & 0 \\
 -i & 0 & 0 \\
\end{array}
\right), \,
S_z=\!\left(
\begin{array}{ccc}
 0 & -i & 0 \\
 i & 0 & 0 \\
 0 & 0 & 0 \\
\end{array}
\right).
\end{align}
The representations S and J are equivalent and related by a unitary transformation
\begin{align}
U^\dagger_{SJ} \,S^i\, U_{SJ} = J^i  
\end{align}
that complies with (\ref{eq:UAB}) and where the matrix $U_{SJ}$ is defined by the expression
\begin{align}
    &U_{SJ} = \frac{1}{\sqrt{2}}\left(
\begin{array}{ccc}
 - 1 & 0 & 1 \\
 -i & 0 & -i \\
 0 & \sqrt{2} & 0 \\
\end{array}
\right).
\end{align}

The transition from the J to the S representation can be understood as a change of basis from $|j \rangle$ to $|e_i \rangle \,(i, j=1,2,3)$, defined by the relation
\begin{equation}
|e_i \rangle = \sum_j |j \rangle \left(U_{SJ}^\dagger \right)_{j i}.   
\end{equation} 
An explicit calculation gives:
\begin{eqnarray}
| e_1 \rangle &=& \frac{1}{\sqrt{2}} (|1,-1 \rangle -   |1,+1 \rangle , \nonumber \\
| e_2 \rangle &=& \frac{i}{\sqrt{2}} (|1,-1 \rangle +   |1,+1 \rangle,  \label{eq:es} \\
| e_3 \rangle &=& |1,0 \rangle . \nonumber
\label{eq:ei}
\end{eqnarray}
We note that these states satisfy the equations $J_x | e_1 \rangle = 0$, $J_y | e_2 \rangle = 0$, and $J_z | e_3 \rangle=0$. 

Any pure state can be represented as a linear combination of the sta-\linebreak tes~$| e_i \rangle$,
\begin{equation}
\varepsilon^1_{*}  | e_1 \rangle +   \varepsilon^2_{*}  | e_2 \rangle +  
\varepsilon^3_{*}  | e_3 \rangle.
\end{equation}
Hence, it is defined by a three-vector $\varepsv_* = \left(\varepsilon^1_{*},\varepsilon^2_{ *},\varepsilon^3_{*}\right)$, where the coefficients $\varepsilon^i_{*}$ are, in general, coefficients. The three-vector $\varepsv_*$ can be generalized to a four-vector through the definition
\begin{equation}
\varepsilon^\mu_* = (0, \varepsv_*) = \left(0, \varepsilon^1_{*},\varepsilon^2_{ *},\varepsilon^3_{*}\right).
\end{equation}
It is also useful to introduce the basis $\epsilon^{\mu}_{r *} \, (r=1,2,3)$,
\begin{align}
    \epsilon^{\mu}_{1*}=(0,1,0,0), \nn \\
    \epsilon^{\mu}_{2*}=(0,0,1,0), \label{eq:Sbasis} \\
    \epsilon^{\mu}_{3*}=(0,0,0,1) \nn,
\end{align}
which corresponds to the linear-polarization states $| e_i \rangle$ defined above by Eq.~(\ref{eq:ei}). In the following, we will often use the relation 
\begin{equation}
\epsilon^{i}_{r*} = \delta^i_r = - \delta_{i r}.
\label{eq:epsir}
\end{equation}

\section{Lorentz-covariant spin density matrix}

We may define the polarization vectors $\epsilon^{\mu}_{r}(p)$ for particles with momentum $p^\mu$ using the canonical boost $L^\mu_{\,\,\nu}(\vv_p)$, which transforms $p^\nu_* = (m,0,0,0)$ into $p^\mu= (E_p, \pvs)$ \CITn{Florkowski:2017dyn},
\begin{align}
p^\mu = L^\mu_{\,\,\nu}(\vv_p) p^\nu_* , \qquad 
\epsilon^{\mu}_{r} = L^\mu_{\,\,\nu}(\vv_p)\epsilon^{\nu}_{r*}.
\label{eq:boostedeps}
\end{align}
In the same way, we define the spin polarization four-vector ${\cal P}^\mu$ and the tensor polarizabilities ${\cal T}^{\mu\nu}$ for particles with momentum $p$,
\begin{align}
{\cal P}^\mu = L^\mu_{\,\,\nu}(\vv_p) {\cal P}^\nu_* , \qquad 
{\cal T}^{\mu\nu} = L^\mu_{\,\,\alpha}(\vv_p)
 L^\nu_{\,\,\beta}(\vv_p) {\cal T}^{\alpha\beta}_*.
\end{align}
Here we assume that  ${\cal P}^0_* = 0$ and ${\cal T}^{00}_* ={\cal T}^{i0}_* = {\cal T}^{0i}_* =0$. This leads to the orthogonality conditions:
\begin{eqnarray}
{\cal P}^\mu p_\mu = {\cal P}^\mu_* p_\mu^* &=& 0, \nonumber \\ 
{\cal T}^{\mu\nu} p_\mu p_\nu = 
{\cal T}^{\mu\nu}_* p_\mu^* p_\nu^* &=& 0,
\nonumber \\
{\cal T}^{\mu\nu} p_\mu  = 
{\cal T}^{\mu\nu}_* p_\mu^* &=& 0.
\label{eq:orth}
\end{eqnarray}

At this point, it is convenient to define the spin density matrix in the Lorentz space, 
\begin{align}
\rho_{\mu\nu}(x,p) = \epsilon^{\,\,r}_{\mu}(p)
\rho^{*S}_{rs}(x,p)
\epsilon^{s}_{\nu}(p),
\label{eq:rhomunu}
\end{align}
where the polarization vectors are given by \EQSTWO{eq:Sbasis}{eq:boostedeps} and satisfy the normalization conditions
\begin{align}
\epsilon^{\,\,r}_{\mu}(p) \epsilon^{\,\,\mu}_{s}(p) = - \delta_{rs}.
\label{eq:epsnorm}
\end{align}
We note that throughout this work we employ the basis of the adjoint representation, in which the polarization vectors $\epsilon^{\,\,r}_{\mu}(p)$ are real. Consequently, no complex conjugation appears in \EQ{eq:epsnorm}. Using \EQ{eq:epsir}, we find that in PRF,
\begin{align}
\rho^*_{ij} = \rho^{* S}_{ij}.
\label{eq:rhoij}
\end{align}
In the frame, in which the particles carry momentum $p$, we find
\begin{align}
 \rho_{\mu\nu}(x,p)    =- \frac{1}{3}  \left[\left(g_{\mu\nu}-\frac{p_\mu p_\nu}{m^2}\right) + \frac{3i  \epsilon_{\mu\nu\lambda\rho} {\cal P}^\lambda(x,p) p^\rho}{2m}+ \sqrt{6}\,{\cal T}_{\mu\nu}(x,p)\right].
\label{eq:rhomunucov} 
\end{align}
It can be easily checked that \EQ{eq:rhomunucov} reduces to \EQ{eq:rhoij} in PRF. It is important to stress that the matrix $\rho^{* S}_{rs}$ is defined in the ``spin'' space, where raising or lowering indices does not affect the sign. In contrast, the elements of the matrix $\rho_{\mu\nu}$ generally depend on the index positions. This is reflected in the traces of $\rho^{* S}$ and $\rho$, which differ by a sign
\begin{align}
    \tr(\rho)=\rho_{\mu\nu}g^{\mu\nu} = \rho_{*ij}g^{ij}= -\sum_{i,j=1}^3\rho_{*ij}\delta^{ij}=-\trthree(\rho^S_{*})=-1.
\end{align}

\section{Conserved currents and Wigner function}

With the goal of constructing spin hydrodynamics that incorporates spin-1 particles, we consider a relativistic gas of particles described by the Proca field. In this case we may use the expressions for the energy-momentum and spin tensors of non-interacting particles derived in \CITn{Weickgenannt:2022jes}:
\begin{align}
    T^{\mu\nu}(x) = \int d^4k \,k^\mu k^\nu \, \tr[\mathcal{W}(x,k)],
\label{eq:Tmn}    
\end{align}
\begin{align}
    S^{\lambda,\mu\nu}(x) = 2 i \int d^4k\, k^\lambda \, \mathcal{W}^{[\mu\nu]}(x,k),
\label{eq:Slmn} 
\end{align}
where $\mathcal{W}^{\mu\nu}$ is the Wigner function. Equations \EQSTWOn{eq:Tmn}{eq:Slmn} were obtained in the KG\footnote{The acronym KG stands for an alternative Klein-Gordon type Lagrangian for massive vector fields.} pseudogauge defined in \CITn{Weickgenannt:2022jes}. They are analogs of the expressions derived for spin-\nicefrac{1}{2}  particles in the GLW pseudogauge~\CITn{Florkowski:2018fap}. The symmetric form of the energy-momentum tensor \EQn{eq:Tmn} implies that the spin tensor \EQSn{eq:Slmn} is conserved. 

In this work, we connect the Wigner function with the spin density matrix using the formula
\begin{align}
\mathcal{W}_{\mu\nu}(x,k) &= - \int dP\, \delta^{(4)}(k-p) f_0(x,p) {\cal Z}(x,p) \, \rho_{\nu\mu}(x,p),
\label{eq:gen_WF}
\end{align}
where $dP$ is the Lorentz-invariant integration measure
\begin{align}
dP = \frac{d^3p}{(2\pi)^3 E_p},   
\label{eq:dP}
\end{align}
with $E_p = \sqrt{m^2 + \pvs^2}$ being the on-mass-shell energy. The function $f_0$ is the spinless  equilibrium  (Maxwell-J\"uttner) distribution function
\begin{align}
f_0(x,p) = \exp\left[\xi(x) - p \cdot \beta(x) \right],
\label{eq:f0}
\end{align}
where $\xi = \mu/T$, with $\mu$ being the chemical potential connected with the conserved charge carried by the vector meson (for example, electric charge), and $T$ being the temperature.~The function $\beta^\mu$ is defined as the ratio of the hydrodynamic flow and the temperature, $\beta^\mu = u^\mu/T$. In \EQ{eq:gen_WF}, the function ${\cal Z}(x,p)$ denotes the normalization of the spin density matrix, used when the original spin density matrix is not normalized to unity. For unpolarized particles, ${\cal Z} = 3$. For simplicity, in the following we set $\mu = 0$ and thus consider a single species of particles whose number is not conserved.

With the Wigner function~\EQn{eq:gen_WF}, \EQ{eq:Tmn} directly gives
\begin{align}
    T^{\mu\nu}(x) = \int dP \,p^\mu p^\nu \,f_0(x,p) {\cal Z}(x,p),
\label{eq:Tmnp}    
\end{align}
while for the spin tensor we obtain
\begin{align}
S^{\lambda,\mu\nu}(x)  
=\frac{1}{m} \int dP\, p^\lambda  f_0(x,p)  {\cal Z}(x,p) \epsilon^{\mu\nu\alpha\beta} {\cal P}_\alpha(x,p) p_\beta .
\label{eq:Slmnp}  
\end{align}
As expected, the spin tensor depends solely on the spin polarization vector ${\cal P}$ and not on the tensor polarizabilities ${\cal T}$.

\section{Local equilibrium spin density}

We have already introduced the concept of thermalization of particle momenta with the help of the function $f_0$. We now turn to the problem of describing the thermal distribution of spin degrees of freedom. In \CITn{Xia:2020tyd} it was argued that the equilibrium spin density matrix takes the form 
\begin{align}
\rho^S_{\rm eq} =  \frac{\exp[\alphav \cdot \bm{S}]}{\trthree(\exp[\alphav \cdot \bm{S}])} 
= {\cal Z}^{-1} \exp[\alphav \cdot \bm{S}],\label{eq:spn_dnsty1}
\end{align}
where $\alphav$ plays a role of the angular velocity. The exponential function appearing in \EQn{eq:spn_dnsty1} can be expanded as a series and rewritten as
\begin{align}
     e^{\alphav\cdot \bm{S}} = 1+{(\hat{\alphav}\cdot \bm{S})} \sinh\alpha + {(\hat{\alphav}\cdot \bm{S})^2}(\cosh\alpha-1),
\end{align}
where we have introduced the notation $\hat{\alphav} = \alphav/|\alphav|$ and $|\alphav|=\alpha$. The normalization in the denominator is
\begin{align}
{\cal Z} = 2 \cosh \alpha +1.
\end{align}

Equation~\EQn{eq:spn_dnsty1} may be used to determine the spin polarization vector and tensor polarizabilities:
\begin{align}
\bm{\mathcal{P}}_* = \frac{2 \sinh (\alpha) }{2 \cosh \alpha +1}  \,\hat{\alphav},
\label{eq:eqbm_spn_pol}
\end{align}
\begin{align}
\mathcal{T}_*^{ij} = \left\{\begin{array}{c}        \frac{\left(3({\hat \alpha}^i)^2\ - 1 \right) \left(\cosh \alpha-1\right)}{\sqrt{6} \, (2 \cosh \alpha +1) } \quad \hbox{for} \quad i=j, \\ \\
\sqrt{\frac{3}{2}}\,\frac{ {\hat \alpha}^i {\hat \alpha}^j \left(\cosh \alpha-1\right)}{ 2 \cosh \alpha +1 } \quad \hbox{for} \quad i\neq j.
\end{array}\right.
\end{align}
It is instructive to rewrite \EQn{eq:eqbm_spn_pol} using the exponential functions
\begin{align}
\bm{\mathcal{P}}_* = \frac{e^\alpha - e^{-\alpha}}{e^\alpha + 1 + e^{-\alpha}}  \,\hat{\alphav}.
\label{eq:eqbm_spn_pol1}
\end{align}
This formula suggests that the factor $e^\alpha$ determines the probability of having spin 1 oriented along the direction $\hat{\alphav}$. An analogous expression for spin \nicefrac{1}{2}  would be
\begin{align}
\bm{\mathcal{P}}^{(\nicefrac{1}{2})}_* = \frac{1}{2} \,\frac{e^{\alpha/2} - e^{-\alpha/2}}{e^{\alpha/2} + e^{-\alpha/2}}  \,\hat{\alphav}.
\label{eq:eqbm_spn_pol12}
\end{align}
Comparing \EQ{eq:eqbm_spn_pol12} to \EQ{eq:eqbm_spn_pol1}, we see that the prefactor $1/2$ associated with the spin of fermions becomes unity when describing the system of bosons.
In fact, the formula \EQn{eq:eqbm_spn_pol12} was obtained in our previous work, in which we established the relation
\begin{align}
\alphav = 2 \av_*,
\label{eq:alphaa}
\end{align}
where $\av_*$ is the spatial part of the four-vector $a^\mu$ defined in the PRF. The four-vector $a^\mu$ is related to the spin polarization tensor $\omega^{\mu\nu}$ through the relation
\begin{align}
    a^\mu=-\frac{1}{2m}\tilde{\omega}^{\mu\nu}p_\nu.
\end{align}
The tensor $\omega^{\mu\nu}$ plays a role of the Lagrange multiplier controlling the conservation of the spin part of the angular momentum. The explicit expression for $\av_*$ reads~\CITn{Florkowski:2017dyn} 
\begin{align}
 \av_\ast &=  -\frac{\bv_*}{2}  = -\frac{1}{2m} \left(  E_{p} \, \bv - \pvs \times \evs - \frac{\pvs \cdot \bv}{E_{p} + m} \pvs \right),
\end{align}
where $\evs$ and $\bv$ are the electric- and magnetic-like components of the tensor $\omega^{\mu\nu}$. The vector $\bv_*$ is the magnetic part of $\omega^{\mu\nu}$ in PRF.

The fact that local-equilibrium spin densities are determined by the local magnetic components of the spin polarization tensor is analogous to the coupling of the local magnetic field to particle magnetic moments. We also note that, in global equilibrium, the spin polarization tensor $\omega^{\mu\nu}$ coincides with thermal vorticity. In this case, the vector $\alphav$ is determined by the vortical properties of the fluid.

\section{Thermodynamic relations and Pauli-Luba\'nski four-vector}

\subsection{Thermodynamics}

With identification \EQn{eq:alphaa}, we find the following forms of the energy-momentum and spin tensors:
\begin{align}
T^{\mu\nu}(x) &= \int dP \, p^\mu p^\nu f_0 \left[2\cosh(2a_*)+1 \right],
\label{eq:Tmna}
\end{align}
\begin{align}
S^{\lambda,\mu\nu}(x) = 2\int dP\, p^\lambda f_0\frac{\sinh (2\sqrt{-a^2})}{m\sqrt{-a^2}}  \epsilon^{\mu\nu\alpha\beta} a_\alpha p_\beta .
\label{eq:Slmna}
\end{align}
Following our previous analyses~\cite{Bhadury:2025boe,Kar:2025qvj}, we introduce the particle current
\begin{align}
\mathcal{N}^\mu=\int dP p^\mu \tr(\mathcal{W}) = \int dP p^\mu f_0 \left[2\cosh(2a_*)+1 \right].
\label{eq:calNma}    
\end{align}
Using \EQn{eq:f0} with the vanishing chemical potential ($\mu=0$) and the derivative
\begin{align}
    d\left(\sqrt{-a^2}\right) = \frac{1}{4 m\sqrt{-a^2}}\epsilon^{\rho\sigma\beta\kappa}a_{\beta}p_{\kappa}d\omega_{\rho\sigma},
\end{align}
we find the thermodynamic relation
\begin{align}
d\mathcal{N}^\mu &=-T^{\mu\lambda}d\beta_{\lambda}+\frac12S^{\mu,\rho\sigma}d\omega_{\rho\sigma}.
\label{eq:dcalN}
\end{align}
Equation~\EQn{eq:dcalN} implies that the energy-momentum tensor and the spin tensor can be obtained by differentiating the particle current $\mathcal{N}^\mu$ with respect to the appropriate Lagrange multipliers, namely
\begin{align}
T^{\mu\lambda} &=-\frac{\partial\mathcal{N}^\mu}{\partial \beta_{\lambda}},
\quad
S^{\mu,\rho\sigma} = \frac{\partial\mathcal{N}^\mu}{\partial \omega_{\rho\sigma}}.
\label{eq:der}
\end{align}
With the conservation laws
\begin{align}
\partial_\mu T^{\mu\lambda} &= 0,
\quad
\partial_\mu S^{\mu,\rho\sigma} = 0,
\label{eq:conlaw}
\end{align}
our description agrees with the structure of the divergence-type \linebreak theory~\cite{Abboud:2025shb,Bhadury:2025wuh}. The form of~\EQ{eq:dcalN} also suggests that one can introduce the entropy current defined by the expression
\begin{align}
S^\mu &= T^{\mu\lambda} \beta_{\lambda} -\frac12 S^{\mu,\rho\sigma} \omega_{\rho\sigma} + \mathcal{N}^\mu.
\label{eq:ent}
\end{align}
Equations~\EQSTWOn{eq:dcalN}{eq:ent} lead to the differential of the entropy current
\begin{align}
dS^\mu &= \beta_{\lambda} dT^{\mu\lambda} - \frac12 \omega_{\rho\sigma} dS^{\mu,\rho\sigma} .
\label{eq:dS}
\end{align}
This expression indicates that the entropy is conserved as a direct consequence of the conservation of the energy, linear momentum, and spin. Following our previous studies, we interpret this case as the state of {\it local thermodynamic equilibrium} for spin-carrying particles. The dynamics of such a state is described by {\it perfect-fluid spin hydrodynamics}.

\subsection{Pauli-Luba\'nski four-vector}

Following earlier studies~\cite{Florkowski:2017dyn}, we define the phase space density of the Pauli-Luba\'nski four-vector by the formula
\begin{align}
E_p\frac{d \Delta \Pi_\mu(x,p) }{d^3p} = -\frac12\epsilon_{\mu\nu\alpha\beta} \Delta\Sigma_{\lambda}(x) E_p\frac{dJ^{\lambda,\nu\alpha}(x,p)}{d^3p}\frac{p^\beta}{m}.
\label{eq:PL1} 
\end{align}
The quantity $\Delta \Sigma_\lambda$ represents a volume element over which particles with momentum $p$ are considered. After integration over momentum and space we obtain the total conserved charges, for example
\begin{align}
    J^{\nu\alpha}_{\rm tot}= \int d\Sigma_{\lambda}(x) \int dP \, J^{\lambda,\nu\alpha}(x,p).
\end{align}
Splitting the angular momentum tensor into orbital and spin parts, \linebreak \mbox{$J^{\lambda,\nu\alpha}(x,p) = L^{\lambda,\nu\alpha}(x,p)+S^{\lambda,\nu\alpha}(x,p)$}, we find that only the spin part contributes to \EQn{eq:PL1} -- the orbital part includes terms proportional to $p^\nu$ and $p^\alpha$, which vanish upon contraction with the Levi-Civita symbol in \EQn{eq:PL1}. Thus, we arrive at the formula
\begin{align}
    E_p\frac{d \Delta \Pi_\mu }{d^3p} = -\frac{1}{2 m^2}\epsilon_{\mu\nu\alpha\beta}  \frac{\Delta\Sigma_{\lambda}(x)}{(2\pi)^3} p^\lambda \mathcal{Z}{f}_0   (x,p)\epsilon^{\nu\alpha\rho\sigma} {\cal P}_\rho p_\sigma p^\beta.
\label{eq:inf_pol}
\end{align}
Similarly, we obtain the number of particles in the volume $\Delta\Sigma_{\lambda}(x)$,
\begin{align}
E_p\frac{d \Delta N }{d^3p}  &= E_p \Delta\Sigma_\lambda \frac{d\mathcal{N}^\lambda}{d^3p}=\frac{1}{(2\pi)^3}\Delta\Sigma_\lambda p^\lambda \mathcal{Z}{f}_0(x,p).
\label{eq:inf_num}
\end{align}
This allows us to find the polarization per particle. Calculating the ratio of Eq. (\ref{eq:inf_pol}) and Eq. (\ref{eq:inf_num}), we obtain
\begin{align}
\frac{d \Delta \Pi_\mu}{d \Delta N} &= -\frac{1}{2m^2}\epsilon_{\mu\nu\alpha\beta}      \epsilon^{\nu\alpha\rho\sigma} {\cal P}_\rho p_\sigma p^\beta
=  {\cal P}_\mu  - (\mathcal{P}\cdot p)p_\mu.
\end{align}
Since $\mathcal{P}\cdot p =0$, see \EQ{eq:orth}, we find
\begin{align}
     \frac{d \Delta \Pi_\mu}{d \Delta N}= {\cal P}_\mu .
\end{align}
This form supports our expression that relates the Wigner function to the spin density matrix~\EQn{eq:gen_WF}.

It is interesting to compare our results with the expressions obtained before for spin-\nicefrac{1}{2}. The formula derived in \CITn{Bhadury:2025boe} gives 
\begin{align}
\frac{d \Delta \Pi_\mu^{(\nicefrac{1}{2})}}{d \Delta N}= \frac{1}{2} \tanh\sqrt{-a^2} \, \frac{a_\mu }{\sqrt{-a^2}},
\label{eq:Pi12}    
\end{align}
which has the following limits for $\sqrt{-a^2} \ll 1$ and $\sqrt{-a^2} \gg 1$, respectively:
\begin{align}
\frac{d \Delta \Pi_\mu^{(\nicefrac{1}{2})}}{d \Delta N}= \frac{1}{2} \, a_\mu , \qquad
\frac{d \Delta \Pi_\mu^{(\nicefrac{1}{2})}}{d \Delta N}= \frac{1}{2}  \, \frac{a_\mu }{\sqrt{-a^2}}.
\label{eq:Pi12smalllarge}   
\end{align}
Combining~\EQSTWOn{eq:Slmnp}{eq:Slmna} we obtain 
\begin{align}
\frac{d \Delta \Pi_\mu}{d \Delta N}=  \frac{2 \sinh (2 \sqrt{-a^2})}{2 \cosh (2 \sqrt{-a^2}) + 1}  \, \frac{a_\mu }{\sqrt{-a^2}},
\label{eq:Pi1}    
\end{align}
with the corresponding limits (again for $\sqrt{-a^2} \ll 1$ and $\sqrt{-a^2} \gg 1$):
\begin{align}
\frac{d \Delta \Pi_\mu}{d \Delta N} = \frac{4}{3} a_\mu, \qquad  
\frac{d \Delta \Pi_\mu}{d \Delta N} =\frac{a_\mu }{\sqrt{-a^2}}. 
\label{eq:Pismalllarge}   
\end{align}
It is instructive to see that for $\sqrt{-a^2} \gg 1$, the quantity $d\Delta \Pi_\mu / d \Delta N$ is bounded by 1/2 and 1, respectively, for spin 1/2 and 1. This indicates the correct normalization of our spin densities. 

\section{T representation and alignment measurements}
\label{sec:alignent}
Let us turn to the discussion of the T representation, which is used in  measurements of the spin alignment. The $T^i$  matrices are explicitly given by
\begin{align}
T_x& = \frac{i}{\sqrt{2}}\left(
\begin{array}{ccc}
 0 & -1 & 0 \\
 1 & 0 & -1 \\
 0 & 1 & 0 \\
\end{array}
\right), \quad
T_y=\left(
\begin{array}{ccc}
 1 & 0 & 0 \\
 0 & 0 & 0 \\
 0 & 0 & -1 \\
\end{array}
\right), \notag\\[0.3em]
&\hspace{1.5cm}T_z= \frac{1}{\sqrt{2}}\left(
\begin{array}{ccc}
 0 & 1 & 0 \\
 1 & 0 & 1 \\
 0 & 1 & 0 \\
\end{array}
\right).
\end{align}
The unitary matrices that connect the J and S representations with the T representation are:
\begin{align}
U_{JT} = \left(
\begin{array}{ccc}
 -\frac{1}{2} & -\frac{1}{\sqrt{2}} & -\frac{1}{2} \\[0.3cm]
 -\frac{i}{\sqrt{2}} & 0 & \frac{i}{\sqrt{2}} \\[0.3cm]
 \frac{1}{2} & -\frac{1}{\sqrt{2}} & \frac{1}{2} \\[0.3cm]
\end{array}
\right),\quad U_{TS} = \left(
\begin{array}{ccc}
 \frac{1}{\sqrt{2}} & 0 & \frac{i}{\sqrt{2}} \\[0.3cm]
 0 & -i & 0 \\[0.3cm]
 \frac{1}{\sqrt{2}} & 0 & -\frac{i}{\sqrt{2}} \\[0.3cm]
\end{array}
\right).\notag
\end{align}
From this we can obtain the spin density matrix in  the T representation. Since $\rho_{rs}^T$ is Hermitian, its elements read:
\begin{align}
    \rho_{11}^T&= \frac{1}{6}\left(2+3 \mathcal{P}_*^2\!+\!\sqrt{6} \mathcal{T}_*^{22}\right),\\[0.3em]
    \rho_{21}^T=   \bar{\rho}_{12}^T&= \frac{3 i {\mathcal{P}_*^1}+3 {\mathcal{P}_*^3} +2 \sqrt{6} ({\mathcal{T}_*^{23}}+i {\mathcal{T}_*^{12}})}{6\sqrt{2}},\\[0.3em]
    \rho_{13}^T=\bar{  \rho}_{31}^T&= -\frac{{2 {\mathcal{T}_*^{11}}+2 i {\mathcal{T}_*^{13}}+{\mathcal{T}_*^{22}}}}{\sqrt{6}}, \\[0.3em]
    \rho_{22}^T&= \frac{1}{3}-\sqrt{\frac{2}{3}} \mathcal{T}_*^{22},\\[0.3em]
    \rho_{23}^T= \bar{\rho}_{32}^T&=\frac{-3 i {\mathcal{P}_*^1}+3 {\mathcal{P}_*^3}-2 \sqrt{6} ({\mathcal{T}_*^{23}}-i{\mathcal{T}_*^{12}})}{6\sqrt{2}},\\[0.3em]
    \rho_{33}^T&= \frac{1}{6}\left(2-3 {\mathcal{P}_*^2}+\sqrt{6} {\mathcal{T}_*^{22}}\right),
\end{align}
where the overbar indicates complex conjugation.

In the spin literature, it is common to label the elements of the spin density matrix $\rho^T$ with indices $-1,0,1$ instead of $1,2,3$  (to underline the connection between the matrix elements and the spin quantum numbers). Therefore, in this section, we adopt the traditional indexing scheme (e.g., $\rho_{0-1}=\rho^T_{12}$, $\rho_{00}=\rho^T_{22}$, $\rho_{10}=\rho^T_{32}$, $\rho_{1-1}=\rho^T_{31}$, etc.) when presenting expressions for experimentally studied observables. 

Of particular importance is the central element of the spin density matrix,  which defines the alignment,
\begin{align}
\mathcal{A}=\rho_{00} - \frac{1}{3}=-\sqrt{\frac{2}{3}} \mathcal{T}_*^{22}.
\end{align}
We observe that $\rho_{00}$ is determined by $\mathcal{T}_*^{22}$, although it is measured in the $T$ representation. In this respect, our approach differs from \CITn{Li:2022vmb}, where the alignment is expressed by $\mathcal{T}_*^{33}$. Furthermore, in contrast to~\CITn{Li:2022vmb}, our analysis indicates that a non-trivial alignment can occur in local equilibrium -- it does not require dissipation. Using the expression
\begin{align}
    \mathcal{T}^{22}_*=\frac{\left(3{{\hat \alpha}}_{2}^{2}\ - 1 \right) \left(\cosh \alpha-1\right)}{\sqrt{6} \, (2 \cosh \alpha +1) },
\end{align}
we find that the alignment is given by the formula
\begin{align}
    \mathcal{A}=-\frac{\left(3{{\hat \alpha}}_{2}^{2}\ - 1 \right) \left(\cosh \alpha-1\right)}{3 \, (2 \cosh \alpha +1) }.
\end{align}
For small values of $\alpha$,
\begin{align}
    \mathcal{A} \approx -\frac{ 1}{18} \left(3 {{\hat b}_{2\,*}}^{2} - 1 \right)|\bv_*|^2.
\end{align}
where we have used the fact that $\alphav=2\av_*=-\bv_*$.
Other experimentally accessible observables include~\CITn{Xia:2020tyd}:
\begin{align}
    -\sqrt{2}[{\rm Re}(\rho_{10})-{\rm Re}(\rho_{0-1})]&=\frac{2\sqrt{2}}{3}\mathcal{T}_*^{23},\\
    \sqrt{2}[{\rm Im}(\rho_{10})-{\rm Im}(\rho_{0-1})]&=-\frac{2\sqrt{2}}{3}\mathcal{T}_*^{12},\\
    -\sqrt{2}{\rm Re}(\rho_{1-1})&=\frac{\mathcal{T}_*^{11}-\mathcal{T}_*^{33}}{\sqrt{3}},\\
    \sqrt{2}{\rm Im}(\rho_{1-1})&=\frac{2\mathcal{T}_*^{13}}{\sqrt{3}}.
\end{align}
We observe that in all these cases, the measured quantities can be expressed by the tensor polarizabilities rather than the spin polarization vector. 

We conclude this section by the remark that knowledge of the spin polarization tensor $\omega^{\mu\nu}$ on the freeze-out hypersurface $\Sigma^\lambda$ allows us to obtain the alignment for particles with a given momentum,
\begin{equation}
{\cal A}(p) = \frac{
\int d\Sigma_\lambda p^\lambda \mathcal{Z}{f}_0(x,p) {\cal A}(x,p)
}{\int d\Sigma_\lambda p^\lambda \mathcal{Z}{f}_0(x,p)}.    
\end{equation}

\section{Summary}

In this work, we have studied several issues related to the description of spin-1 particles in heavy-ion collisions. We have demonstrated that the adjoint representation is particularly convenient for analyzing spin observables and calculating the energy-momentum and spin tensors. The local equilibrium spin density matrix was introduced and its advantages emphasized: it leads to thermodynamically consistent relations and allows to construct a unified description of spin-1 and spin-$\nicefrac{1}{2}$ particles. The last property may soon allow a direct check of wether the spin polarization data for hyperons and vector-mesons are driven by the same physical mechanism. We have also shown that the KG definition of the spin tensor is consistent with the definition of the Pauli-Lubański four-vector, and that the obtained equations of spin hydrodynamics have the structure of a divergence-type theory.

\section*{Acknowledgements}
This work was supported in part by the National Science Centre (NCN), Poland, Grant No. 2022/47/B/ST2/01372.

\bibliographystyle{apsrev4-1}
\bibliography{references.bib}
\end{document}